\documentstyle[epsfig]{elsart}

\begin{document}
\hspace*{\fill}%
\begin{minipage}{7cm}
\flushright
UMIST/Phys/TP/96-3\\
nucl-th/9607024
\end{minipage}

\begin{frontmatter}
\title{Translationally invariant treatment of pair correlations in nuclei: I.
Spin and isospin dependent correlations}

\author{R. Guardiola, P. I. Moliner}
\address{Dpto. de F\'{\i}sica At\'omica, Molecular y Nuclear,
Universitat de Val\`encia,
Avda. Dr. Moliner 50, E-46100 Burjassot, Spain  }

\author{J. Navarro}
\address{IFIC (Centre Mixt CSIC -- Universitat de Val\`encia),
Avda. Dr. Moliner 50, E-46100 Burjassot, Spain  }

\author{R.F. Bishop, A. Puente, Niels R. Walet\thanksref{emailNRW}}
\thanks[emailNRW]{electronic address: Niels.Walet@umist.ac.uk}

\address{Department of Physics, UMIST, P.O. Box 88,
Manchester M60 1QD, U.K.}

\begin{abstract}
We study the extension of our translationally invariant treatment of few-body
nuclear systems to 
heavier nuclei. At the same time we also
introduce state-dependent correlation operators. Our techniques are tailored to
those nuclei that can be dealt with in $LS$ coupling, which includes all 
nuclei up to the shell closure at $A=40$.
We study mainly $p$-shell nuclei in this paper.
A detailed comparison with other microscopic many-body approaches is made,
using a variety of schematic nuclear interactions. It is shown that our
methodology produces very good energies, and 
presumably also wave functions,
for medium mass nuclei.
\end{abstract}
\end{frontmatter}

\section{Introduction}
During the last few years there has been an impressive improvement in
the methodology used to perform fully microscopic calculations for
the ground-state properties of
nuclear systems.  Most progress has been made for
light nuclei, particularly the systems of three and four nucleons,
where quite different methods, including Monte Carlo methods \cite{carlson},
Faddeev and Faddeev-Yakubovski methods \cite{gloeckle},
hyperspherical harmonic expansions \cite{rosati}, and
translationally invariant configuration interaction methods 
\cite{us}, have been pushed to a high level of reliability. 

Ideally one would like to extend as many of the methods as possible to
heavier systems.  Unfortunately, however, some of these methods are specially
designed for very light systems, and 
cannot easily be extended to
heavier systems.  The 
Green's function 
Monte Carlo methods have been extended to 
systems of up to six nucleons, but it is unlikely that such methods will be
able to tackle systems 
with  more than a few additional nucleons in the
foreseeable future. The hyperspherical harmonic method is also
limited in the number of nucleons it can treat, even though the Pisa
group has started an ambitious project for $p$-shell nuclei.
Arias de Saavedra {\em et al.}\/ \cite{ACFF96} have recently investigated
the application of the Fermi hypernetted chain (FHNC) method to
heavier nuclei, with some apparent success. Another promising approach is
one based on a stochastic variational principle \cite{VS}, but this method 
is also limited to relatively light nuclei.
We should also mention the variational Monte Carlo
calculations of $^{16}$O \cite{kalos,pandha} which have considered
realistic interactions. These impressive calculations, however,
have used a multiplicative cluster expansion to deal with the
operatorial structure of the trial wave function.

In this paper we shall discuss another approach, which was pioneered
by us some years ago \cite{us}. This translationally invariant
configuration interaction (TICI) method was inspired by the coupled
cluster (CC) method. In its lowest-order implementation (CC2) 
the coupled cluster method
consists of using wave functions obtained by acting on a reference
state with the exponentiated 
one- and two-body correlation operators, 
$\exp[S_1+S_2]$. The CC method is most naturally formulated in the
occupation-number representation. However, as was to be expected, we 
found very slow convergence with
respect to the cut-off on the single-particle basis in this
representation. This led us to consider a linearised form of the CC2
approach, namely the TICI2 method, which also includes pair correlation 
effects but which can easily be converted to coordinate representation.
It has the trial wave function
\begin{equation}
\label{eq:wf0}
|\Psi\rangle = \sum_{i<j} f(r_{ij}) |\Psi_0\rangle.
\end{equation}
In order to remove spurious effects due to the
centre-of-mass motion the uncorrelated wave function $|\Psi_0\rangle$ must be 
a harmonic-oscillator shell-model function. In contrast to the standard CC2
method, the TICI2 method is variational, and leads to an upper
bound to the ground-state energy.

The simple form of eq.\ (\ref{eq:wf0}) for the wave function has provided 
very good 
estimates of the ground-state energy of $^{4}$He for simple Wigner forces 
\cite{us}.  When we revert
 to occupation-number representation,
we find that the Ansatz (\ref{eq:wf0}) 
corresponds to the most general shell-model excitation operators containing 
both one particle-one hole and two particle-two hole components.  These 
components have to be appropriately related so as to ensure the 
translational invariance.

One obvious improvement of this Ansatz is the introduction of
three-body correlations, of the form
\begin{equation}
\label{wf1}
|\Psi\rangle =\left( \sum_{i<j} f(r_{ij})+
\sum_{i<j<k} g({r}_{ij}, {r}_{ik},{r}_{jk} )\right)
 |\Psi_0\rangle.
\end{equation}
Our calculations using this form of trial wave function, 
with a simple model interaction in $^4$He,
produced almost the exact energy for the ground state \cite{us}.  
As one can see, we have a systematic way of improving the wave function, 
by adding terms of physical relevance. One may expect that, in medium light
nuclei, one must include
only up to three- and maybe four-body
correlations, in order to get a precise determination of the energy
and the structure of the system.

On the other hand, one must also deal with the complex operatorial
structure of realistic nucleon-nucleon (NN) interactions, which in
turn requires an additional operatorial structure for the correlation
operator.  At the same time, one also needs to find a way to extend
the method to deal with nuclei with a larger number of particles. As
is also true for some (but not all) of the other approaches cited above, this 
extension is trivial, but the underlying computational complexity
requires an appreciable  effort. We should recall the obvious fact that
the expectation value of a two-particle operator, like the NN
potential, between states of the type of eq.\ (\ref{eq:wf0}) requires
the evaluation of the expectation value of up to six-body
operators. These calculations are large but not prohibitively so.

The dual purpose of this work is to show how this kind of
computation may be carried out in practice, and also to investigate 
the importance of the operatorial structure of the pair correlation.
In this paper we are considering simple interactions of the V4 form 
(central, but spin- and isospin-dependent). We also consider 
a two-particle correlation operator of the same
 structure, i.e., with the spin- and isospin-dependent form
\begin{equation}
\label{eq:pairc}
f_{ij} = f_c(r_{ij}) + f_\sigma(r_{ij}) (\vec{\sigma}_i \cdot\vec{\sigma}_j) +
f_\tau(r_{ij}) (\vec{\tau}_i\cdot \vec{\tau}_j) +
f_{\sigma \tau}(r_{ij}) 
(\vec{\sigma}_i \cdot\vec{\sigma}_j)(\vec{\tau}_i\cdot \vec{\tau}_j).
\end{equation}

In sec.\ \ref{sec:method} we discuss the
computational problems associated with this form of correlation operator 
and our way of tackling them. Our scheme
goes beyond the typical shell-model methodology, which
basically deals with two-body operators. 
In its current form our calculations are limited to selected nuclei,
specifically to those systems which may be
described or approximated by spin-isospin saturated wave
functions. At the cost of losing the rotational invariance, one
may basically deal with $A=4n$ nuclei, where $n$ is an integer. For
some systems (like, e.g., $^{12}$C or $^{8}$Be) it is actually
better to break the rotational symmetry in favour of a
deformed structure. In these nuclei the energy of the deformed state
is much lower than that built on a spherical state
of the {\em jj} coupling scheme. A further restriction
on the nuclei we can deal with originates from the fact that our method
makes intricate use of $LS$-coupling. This means that our method will
not perform well beyond the shell closure at $^{40}$Ca.

Having discussed our methodology in  some considerable detail, we then turn to
a comparison of results in sec.\ \ref{sec:results}.
Our main goal is to gain some insight into both the power of our 
method and any associated shortcomings.
 We make a detailed comparison with other many-body methods and
exact results where available. Finally we draw some conclusions and give
an outlook in sec.\ \ref{sec:DisOut}.

\section{The evaluation of matrix elements \label{sec:method}}

In this section we shall show how to calculate matrix elements of a
V4-type potential, when the correlations are of the form given in
eq.\ (\ref{eq:pairc}). 
We present all
reductions of the required $A$-body matrix elements by identifying
the minimal set of integrals that we are required to calculate.
The reduction of spin and isospin
degrees of freedom is also discussed, for the case of saturated
systems.

The discussion has been limited to the case of the expectation
value of the potential, which is the most complex quantity we need to 
evaluate. A similar, but much less complicated, analysis
can be carried out for the kinetic energy
operator as well as for the computation of the norm
of the pair-correlated state.

\subsection{Diagrammatic analysis}
Let $|\Psi_0\rangle$ represent an independent-particle harmonic-oscillator 
shell-model wave function for the nucleus under
consideration.  For the case of the spin-isospin saturated systems in which we
are interested, this uncorrelated state may be represented by a
{\em single}\/ Slater determinant, which is hence fully characterised by the
enumeration of the occupied single-particle states. Moreover, given
that we are dealing with spin- and isospin-saturated systems, it
is only necessary to specify the harmonic oscillator states occupied,
and we can ignore spin and isospin for the characterisation of the states.
 For example, the uncorrelated state
for $^{16}$O requires the four single-particle states $0s$,
$0p_{+1}$, $0p_{0}$ and $0p_{-1}$, where the subscript refers to the
magnetic quantum number.  
We can transform to a Cartesian basis where the occupied single-particle
states are represented by the three harmonic oscillator
quantum numbers $(n_x,n_y,n_z)$. The reference state for $^{16}$O
in this basis consists of Slater determinant formed from the four 
occupied states $(0,0,0)$, $(1,0,0)$, $(0,1,0)$ and $(0,0,1)$.

Consider now the expectation value of the potential energy operator,
given by
\begin{equation}
\label{potential_1}
\langle V \rangle =
\langle \Psi_0 | \sum_{i<j} f_L(ij) 
\sum _{k<l} V(kl) \sum _{m<n} f_R(mn) | \Psi_0 \rangle,
\end{equation}
where the subscripts $L$ and $R$ identify (for later ease of discussion)
the correlation operators at the
left hand side and the right hand side of the potential
operator,  respectively.
The indices in parentheses label the particles involved
in the corresponding  correlation or  potential operator.
When we expand the sums in eq.\ (\ref{potential_1}) there appears a
total of 
${\setlength{\arraycolsep}{2pt}
                     \renewcommand{\arraystretch}{0.65}
                        \left( \begin{array}{c} {\scriptstyle A} \\
                                                {\scriptstyle 2}
                                                \end{array} \right)}^3$
terms.  One way to classify the individual terms
is by the number $N$ of particles involved.  This number ranges from 2
(e.g., a term $f_L(12) V(12) f_R(12)$) up to six (e.g., a term
$f_L(12) V(34) f_R(56)$). Moreover, in order to reduce the
$A$-particle matrix element to an $N$-particle matrix element it is
convenient to decompose the  operator of eq.\ (\ref{potential_1}) into
a sum of {\em symmetric} $N$-body operators. It is very simple to make
a diagrammatic analysis of these operators by assigning to all $f$'s
and $V$'s a line connecting the points specified in their arguments \cite{us}.
The topologically distinct diagrams 
constitute a natural basis for the 
expansion of the matrix elements.

With some patience one then obtains the decomposition
\begin{equation}
\label{potential_2}
\sum_{i<j} f_L(ij) 
\sum _{k<l} V(kl) \sum _{m<n} f_R(mn) =
\sum_{d=1}^{16} \sum_{i_1<i_2<\dots<i_{N_d}}{\cal O}^{d}
_{i_1 i_2 \dots i_{N_d}}
\end{equation}
where the  symmetric operators ${\cal O}^d$ are defined in table 
\ref{diagrams}.
To clarify the meaning of this table, let us evaluate, for example,
 the three body part of 
$f_LVf_R$. According to the table it can be expanded in the diagrams 
2 till 5 as 
\begin{eqnarray*}
\left({f_LVf_R}\right)_{i_1i_2i_3} & = &
\sum_{P} 
[ f_L(i_1i_2) V(i_1i_2) f_R(i_1i_3) +
f_L(i_1i_2) V(i_1i_3) f_R(i_1i_3) +  \\
& & 
 f_L(i_1i_2) V(i_1i_3) f_R(i_1i_2) +
f_L(i_1i_2) V(i_1i_3) f_R(i_2i_3) ]
\end{eqnarray*}
where the sum over permutations $P$ includes the six
permutations of the triplet $(i_1i_2i_3)$. In other cases  there may be
restrictions  on the permutations,
as mentioned in the table.
The number of permutations $P_d$ of a given 
diagram under these restriction is listed in the last column of the table.
This allows the enumeration  of the diagrams to be checked, since they 
must satisfy the equality
\[
\left( \begin{array}{c} A \\2\end{array} \right)^3
 = 
\sum_{d=1}^{16} P_d 
\left( \begin{array}{c} A \\N_d\end{array} \right)
\]
where $d$ refers to a given diagram, and $N_d$ is the
number of particles involved in that diagram.

\begin{table}[htb]
\caption{The topologically distinct diagrams for potential
matrix elements. The column labelled $d$ gives a unique label for each 
diagram, $N_d$ specifies the number of particles involved. 
The column labelled $f_L$ gives the particle labels ($ij$) 
of the left-hand side
correlation function, $V$ those ($kl$) of the potential and $f_R$ those ($mn$) 
of the right-hand side correlation.
The diagrams
must be symmetrised with respect to the  particle
labels $i_1\ldots i_{N_d}$ involved, but with some restrictions to 
avoid overcounting. These 
restrictions are given by the order in which
the labels appear on the functions $f_{L,R}$ and $V$. The
number of permutations $P_d$ under these restrictions is given in the 
last column.}
\label{diagrams}
\begin{center}
\vspace*{0.3cm}
\begin{tabular}{lllllll}
\hline
$d$ & $N_d$ & $f_L$ & $V$ & $ f_R$ & Restrictions& $P_d$ \\
 &  & $ij$ & $kl$ & $mn$ & on Permutations &  \\
\hline
1  &2         & $i_1i_2$      & $i_1 i_2$     & $i_1 i_2$ & $i<j$& 1        \\
\hline
2  &3         & $i_1 i_2$       & $i_1 i_2$     & $i_1i_3$     &
None & 6        \\
3  &         & $i_1 i_2$   & $i_1 i_3$     & $i_1i_3$  & None& 6        \\
4  &         & $i_1 i_2$   & $i_1 i_3$     & $i_1i_2$  & None& 6        \\
5  &         & $i_1 i_2$   & $i_1 i_3$     & $i_2i_3$  & None& 6        \\
\hline
6  &4         & $i_1i_2$   & $i_1i_2$     & $i_3i_4$       &
$i<j, \ m<n$& 6        \\
7  &         & $i_1i_2$    & $i_3i_4$     & $i_3i_4$       &
$i<j, \ m<n$& 6        \\
8  &         & $i_1i_2$    & $i_3i_4$     & $i_1i_2$      &
$i<j, \ k<l$& 6        \\
9  &         & $i_1i_2$  & $i_1i_3$& $i_3i_4$  & None& 24 \\
10 &         & $i_1i_2$  & $i_1i_3$& $i_1i_4$  & None& 24 \\
11 &         & $i_1i_2$  & $i_3i_4$& $i_1i_3$  & None& 24 \\
12 &         & $i_1i_2$  & $i_2i_3$& $i_1i_4$  & None& 24 \\
\hline
13  &5       & $i_1i_2$  & $i_3i_4$& $i_1i_5$  & $k<l$&  60 \\
14  &        & $i_1i_2$  & $i_1i_3$& $i_4i_5$  & $m<n$&  60 \\
15  &        & $i_1i_2$  & $i_3i_4$& $i_3i_5$  & $i<j$&  60 \\
\hline
16  &6        & $i_1i_2$  & $i_3i_4$& $i_5i_6$  & 
$i<j, \ k<l , \ m<n$&  90 \\
\hline
\end{tabular}
\end{center}
\end{table}

After this classification, the expectation value of a given
diagram between Slater determinantal wave functions is reduced to
\begin{eqnarray}
\lefteqn{
\langle \Psi_0 | \sum_{i_1< \dots< i_{N_d}} 
{\cal O}^d_{i_1 \dots i_{N_d}} | \Psi_0 \rangle }  \nonumber \\ 
&= &
\frac{1}{N_d!} 
\sum_{i_1 \dots i_{N_d} = 1}^{A}
\langle
\psi_{i_1}(1) \dots \psi_{i_{N_d}}(N_d)|
{\cal O}^d_{1 \dots N_d}
| \sum_P \epsilon_P
\psi_{Pi_1}(1) \dots \psi_{Pi_{N_d}}(N_d) \rangle. \label{potential_3} 
\end{eqnarray}
This equation requires further explanation. The single-particle
states are represented by the wave functions $\psi_i(j)$, which also 
includes a spinor and an isospinor. Here the subscript $i$ refers 
to the single particle state, and the index in parenthesis specifies the 
particle involved.
The right-hand side of the equation also includes a sum over permutations
{\em of the single-particle state labels}, without any restriction. 
Note that the sum over single particle states extends over all occupied
orbitals.

A very important reduction in the computation is related to a further
symmetry of eq.\ (\ref{potential_3}), and refers to the sum of
permutations (not explicitly written in this equation) which enter
into the definition of ${\cal O}^d$. It turns out that each of  these
permutations gives the same contribution to the matrix element, so that
it is not necessary to evaluate all permutations separately. It is
sufficient to consider one of the diagrams contributing to ${\cal
O}^d$, and multiply the resulting expectation value by the number of
permutations of the diagram 
given in the last column of table 1.

\subsection{The spin and isospin degrees of freedom}

To evaluate a given matrix element, eq.\ (\ref{potential_3}), one
must sum over all occupied single-particle states. In the
case of spin-isospin saturated systems, the index
labelling the single-particle states may be split into three pieces,
\[
i \equiv (i^s,i^\sigma,i^\tau)
\]
where $s$, $\sigma$ and $\tau$ refer to space, spin and isospin,
respectively. The sum over all occupied states is
equivalent to a sum over harmonic-oscillator states, 
together with a replacement of all products of spin and isospin operators
by their respective traces.

The computation of the trace requires a further reduction, since
it interferes with the calculation of permutations. Each
of the permutations appearing in the $N$-particle state
of eq.\ (\ref{potential_3}) can be split into three independent 
permutations, as can be seen from the pair-wise exchange
\[
 (ij) \equiv (i^s j^s) (i^\sigma j^\sigma) (i^\tau j^\tau).
\]
The pairwise permutations referring to spin and isospin indices
can be implemented by having familiar spin and isospin exchange
operators, $P^\sigma_{ij}$ and $P^\tau_{ij}$ act on the wave function. 
Thus the spin and isospin
part of the permutation $P$, $P^\sigma$ and $P^\tau$, can be written
as products of exchange operators, which can be included
inside the trace over spin and isospin.

We thus obtain the matrix element
\begin{eqnarray}
\lefteqn{
\label{trace}
\langle \Psi_0 | f_L(ij) V(kl) f_R(mn) | \Psi_0 \rangle }
\nonumber\\&=& 
\sum_{pqr}
\sum_{i^s_1 \dots i^s_N=1}^{A/4}
\sum_{P} \epsilon_P
{\mathrm{Tr}}\ (\Theta^p_{ij} \Theta^q_{kl} \Theta^r_{mn} 
P^\sigma_{i^\sigma_1\ldots i^\sigma_N}
 P^\tau_{i^\tau_1\ldots i^\tau_N})
 \\
& &
\langle
\phi_{i^s_1}(1) \dots \phi_{i^s_N}(N) |
f_L^p(r_{ij}) V^q(r_{kl}) f^r_R(r_{mn})
|
\phi_{Pi^s_1}(1) \dots \phi_{Pi^s_N}(N) \rangle.
\nonumber 
\end{eqnarray}
Here the $\phi$'s are the spatial (harmonic oscillator) parts of the 
single-particle wave functions $\psi$ in eq.\ (\ref{potential_3}). Since
all spin and isospin sub-states are filled, each  harmonic oscillator
state can be occupied at most four times.
Both the left and right correlation operators  have been expanded in a
basis of operators,
\begin{equation}
 f(ij) = \sum_k f^k(r_{ij}) \Theta^k(ij),
\label{eq:fTheta}
\end{equation}
with a similar expansion for the potential operator.
A quite convenient basis for a V4 form is the set of spin and isospin
exchange operators,
\begin{equation}
\Theta^1(ij) = 1,\;
\Theta^2(ij) = P^\sigma_{ij},\;
\Theta^3(ij) = P^\tau_{ij},\;
\Theta^4(ij) = P^\sigma_{ij}P^\tau_{ij}.
\label{eq:defTheta}
\end{equation}
 Since the spin and isospin permutation
operators can also be written as product of exchange operators, we now
have to evaluate the trace of a long string of exchange operators.

\subsection{The computation of traces}
For the case of a V4 set of operators, both for
correlations and interaction, the calculation of the traces is 
rather simple. 
For an $N$-particle subsystem, a given spin (or isospin) state
may be represented by a column vector of $2^N$ components, where
a given component defines all orientations of the $N$ spins. A
simple representation is obtained by assigning
a binary sequence to a given set $ [ \sigma_1, \sigma_2, \dots , \sigma_N]$ 
of spins:
\begin{equation}
 {\cal N} = \sigma_1 + 2 \sigma_2 + 4 \sigma_3 
\dots + 2^{(N-1)} \sigma_N
\end{equation}
where we use $\sigma_i=0$ for spin up, and $\sigma_i=1$
for spin down. 

With this notation, a spin exchange operator is represented by
a $2^N \times 2^N$ matrix with only one non-zero
element per row and column. This facilitates the storage of matrices, the
multiplication of exchange operators and the computation of the trace.

We have to evaluate a large number of traces, since 
we have  four components for the left correlation, the
potential and the right correlation, as well as the 720 permutations
of a six-particle subsystem, leading to a total of $720\times  4^3=46080$.
With the representation introduced above it is however rather
straightforward -- and very efficient -- 
to write a computer program that evaluates all these traces.

\subsection{Spatial integrals}

In order to make the calculation manageable we expand the correlation
functions in a set of Gaussians, 
\begin{equation}
f^p(r) = \sum_i c^p_i \exp(-\beta_i r^2), \label{eq:expandbeta}
\end{equation}
a technique that has been proven to give an excellent representation of 
the correlation 
function $f$ if we include negative as well as positive values of $\beta$
\cite{us}. This is well matched to the use of
harmonic oscillator single-particle states, since the required space
integrals can be shown to be of the form of the product of an exponential 
of a positive-definite quadratic form with a polynomial.

If the potential is a combination of Gaussians, like the S3
interaction of Afnan and Tang \cite{S3}, it may also be absorbed
into the quadratic form in the exponential. Then, the required
integrals can be computed by using a recurrence relation
\cite{guardiola}.

For other algebraic forms, we can follow two routes. One is to fit a
set of Gaussians to the potential, and apply the technique discussed above.
We choose to follow another route
\cite{guardiola}, where we perform all integrals apart from those 
involving  the
coordinates in the potential by the technique sketched above. The
remaining one-dimensional integral may then be computed by means of a
suitable numerical method.

\subsection{Solution}

Since the different components in the decomposition of eq.\
(\ref{eq:expandbeta}) of the correlation function are not orthogonal,
the expectation value of the energy can be given in the form
\begin{equation}
E(\{c_i^p\}) = \frac
{\displaystyle  \sum _{(ip), (jq)}
c_i^{p*} {\cal H}^{pq}_{ij} c^q_j} 
{\displaystyle  \sum _{(ip), (jq)}
c_i^{p*} {\cal N}^{pq}_{ij} c^q_j}
\end{equation}
where $c^p_i$ is the coefficient multiplying the $i$th Gaussian in the 
correlation function attached to operator $p$. 
If we now optimise with respect to $c_i^{p*}$ we end up with a generalised
eigenvalue problem
\begin{equation}
\sum_{(jq)}
 {\cal H}^{pq}_{ij} c^q_j =
  \lambda \sum_{(jq)} {\cal N}^{pq}_{ij} c^q_j. \label{eq:genev}
\end{equation}
The lowest eigenvalue $\lambda$ gives the approximation for the
ground-state energy.

\subsection{The helium case}

The four-nucleon system deserves a special treatment, since the
decomposition of the two-particle correlation introduced in
eq.\ (\ref{eq:fTheta}) is over-complete for this special nucleus. This
fact is related to the special form of the uncorrelated state, which
is fully space-symmetric, so that the action of the operators tied to
the correlation function is strongly simplified.  In particular, the
action of the spin-isospin exchange operator $P^\sigma_{ij}
P^\tau_{ij}$ is equivalent to $-1$, and the action of $P^\sigma_{ij}$
is the same as the action of $-P^\tau_{ij}$.  Clearly, one should
limit the general form of the correlation so as not to create
duplicate basis states, since this will give rise to problems when
solving the generalised eigenvalue problem of eq.\ (\ref{eq:genev}).
Specifically, the state-dependent correlation of the V4 type for
$^4$He should include only the central scalar (Wigner) and
spin-exchange (Bartlett) pieces. Equivalently, one may include two
separate correlation functions, one for the singlet channels and one
for the triplet channels.

\section{Results} \label{sec:results}

\begin{table}[htb]
\caption{Energies of $p$-shell nuclei for the five interactions
considered, B1, S3, MS3, MT I/III and  MT V.
The column labelled SI corresponds to a purely scalar (state-independent) 
pair correlation, and the column labelled SD
corresponds to a (state-dependent) pair correlation of V4 type.}
\label{table2}
\begin{center}
\renewcommand{\arraystretch}{1.2}
\begin{tabular}{|llrr|}
\hline
\multicolumn{4}{|c|}{B1 interaction}\\
\hline
Nucleus & $\alpha$ (fm$^{-1}$) & $E_{\mathrm{SI}}$ (MeV) &
$E_{\mathrm{SD}}$ (MeV)\\
\hline
$^4$He & 0.729 & -37.86 & -37.86\\
$^8$Be & 0.595 & -49.18 & -61.30\\
$^{12}$C & 0.595 & -84.91 & -103.93 \\
$^{16}$O & 0.602 & -145.94 & -167.30 \\
\hline
\multicolumn{4}{|c|}{S3 interaction}\\
\hline
$^4$He & 0.717 & -25.41 & -28.19\\
$^8$Be & 0.633 & -34.15 & -44.26\\
$^{12}$C & 0.671 & -71.14 & -87.78 \\
$^{16}$O & 0.707 &-141.64 & -164.88 \\
\hline
\multicolumn{4}{|c|}{MS3 interaction}\\
\hline
$^4$He & 0.713 & -25.41 & -27.99\\
$^8$Be & 0.582 & -26.26 & -37.30\\
$^{12}$C & 0.588 & -46.22 & -62.99 \\
$^{16}$O & 0.596 &-85.56 & -105.64 \\
\hline
\multicolumn{4}{|c|}{MT I/III interaction}\\
\hline
$^4$He & 0.741 & -29.45 & -30.81\\
$^8$Be & 0.660 & -46.67 & -52.67\\
$^{12}$C & 0.698 & -99.05 & -109.04 \\
$^{16}$O & 0.744 &-194.10 & -207.52 \\
\hline
\multicolumn{4}{|c|}{MT V interaction}\\
\hline
$^4$He & 0.741 & -29.45 & -29.45\\
$^8$Be & 0.869 & -129.25& -130.23\\
$^{12}$C & 0.997 & -425.99 & -429.44 \\
$^{16}$O & 1.078 &-966.65 & -973.67 \\
\hline
\end{tabular}
\end{center}
\end{table}

Using our technology we have set out to calculate the binding energy
of $^4$He, $^8$Be, $^{12}$C and $^{16}$O. The results are presented in
table \ref{table2}.  In order to allow for the best possible
comparison with existing results in the literature we have
investigated the Brink-Boeker B1 potential \cite{BrinkBoeker}, 
the Afnan-Tang S3 potential \cite{S3}, 
the modified S3 (MS3) potential \cite{GFMP}, 
and the Malfliet-Tjon MT I/III and MT V potentials \cite{MT}. 
We have optimised 
the value of $\alpha$, the inverse 
length-scale of the harmonic oscillator, for the case of the
state-independent (SI), purely central (scalar) correlations, 
i.e., where we restrict
the expansion of eq.\ (\ref{eq:fTheta}) to the $k=1$ term only.
We then performed a
calculation with the full state-dependent (SD) V4 type correlation operator,
keeping the value of $\alpha$
fixed.  The difference between the results from the use of a central
state-independent (SI) and the use of a state-dependent (SD) correlation 
operator is thus probably slightly underestimated. Our calculations
for $^8$Be and $^{12}$C were performed using a 
non-spherical  reference
state (in Cartesian notation)
\begin{equation}
\left|\Psi_0\right\rangle=\left| (0,0,0)^4(0,0,1)^4 \right\rangle
\end{equation}
for  $^8$Be and a similar state,
\begin{equation}
\left|\Psi_0\right\rangle=\left| (0,0,0)^4(1,0,0)^4(0,1,0)^4 \right\rangle
\end{equation}
for $^{12}$C. We recall that we can only use states of this form since we work 
so as to have spin and isospin saturation, within an {\em LS}
coupling scheme. We have broken the rotation symmetry in these cases,
and we no longer get 
a strict estimate for the ground-state energy, but for the slightly higher
intrinsic energy -- a weighted average of the energies of the $J=0,2,4$ 
states comprising the ground-state band. Nevertheless, such an estimate is 
preferable to one using a spherical wave function
(e.g., $\left|(0s_{1/2})^4(0p_{3/2})^8\right\rangle$ for the ground state of 
$^{12}$C). 
This can already be seen from a naive calculation of the $^{12}$C nucleus
using harmonic 
oscillator states and the B1 potential, without correlations.
 The energies in the spherical state
$\left|(0s_{1/2})^4(0p_{3/2})^8\right\rangle$ 
and in the 
non-spherical state $\left| (0,0,0)^4(1,0,0)^4(0,1,0)^4 \right\rangle$ 
are -43.11 MeV and -59.04 MeV, respectively, while  a shell-model 
(or configuration interaction) diagonalisation of the Hamiltonian in the 
$0p$-shell gives -62.07 MeV. Thus an uncorrelated estimate based on the 
deformed state appears to be almost optimal.
Note that the energy may be lowered still further 
by using a deformed single particle basis,
i.e., by using different harmonic oscillator
parameters along the different axes.

The effects of improving the correlations are very interesting. Note
that for the B1 and MT V interactions the introduction of SD-type
correlations does not make a difference for $^4$He. This is due to the
fact that these interactions contain only purely central scalar
(Wigner) and space-exchange (Majorana) terms, and hence do not couple
to the spin-dependent piece of the correlation operator in $^4$He.
Even though there is no spin-dependent part to these potentials,
spin-dependent correlations do play a r\^ole for all heavier nuclei,
so long as the uncorrelated state is not fully space-symmetric.  In
the B1 case the difference is over 20 MeV for $^{16}$O. The effect is
smaller for the MT V potential, but this interaction is very
pathological -- it does not saturate nuclear matter.  For the S3, MS3,
and MT I/III interactions we find that the effect of the additional
correlations is in the 10 to 20 MeV range for $^{16}$O, an important
difference.

\begin{figure}[htb]
\epsfxsize=13cm
\centerline{\epsffile{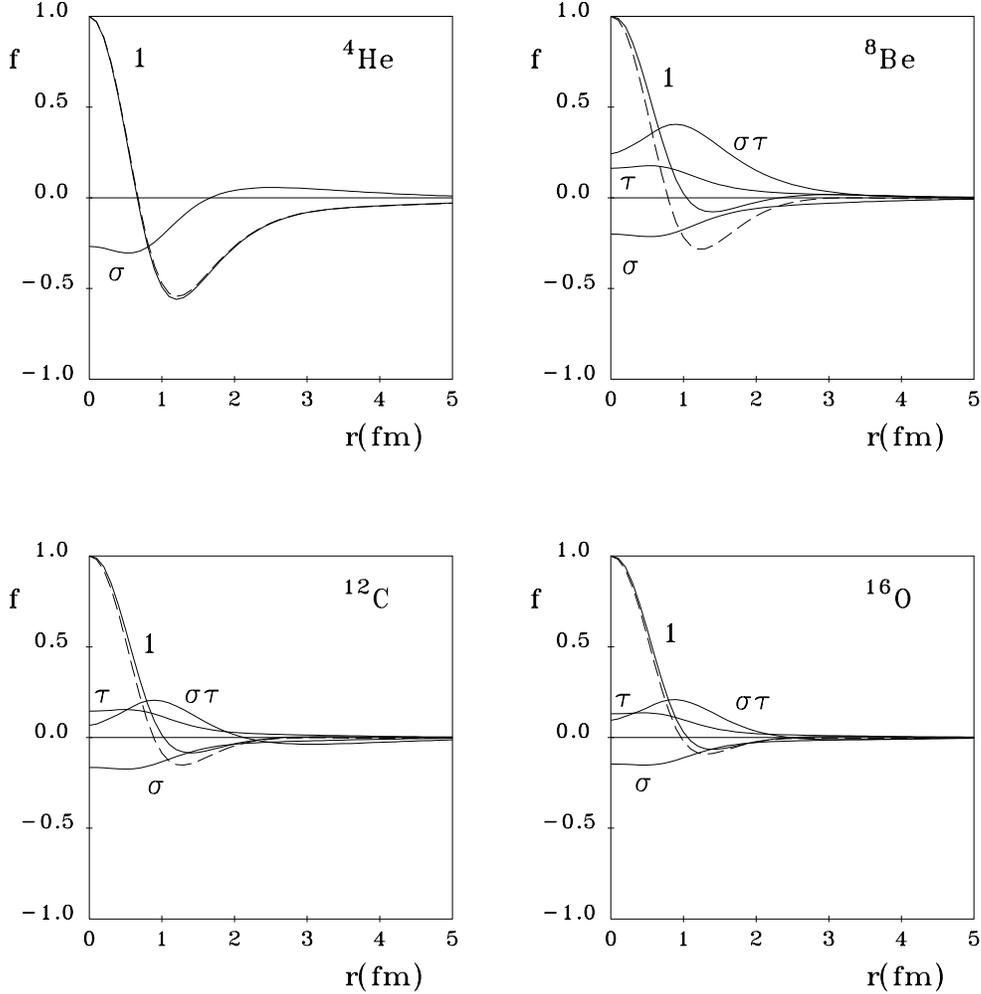}}
\caption{The calculated correlation functions in the case
of the MS3  potential. 
Dashed lines correspond to purely central scalar SI correlations 
and solid lines to SD correlations of V4 type. In the latter case the labels
$1$, $\sigma$, $\tau$ 
and $\sigma\tau$ correspond to the terms $k=1,2,3$
and $4$ in the parametrisation of eqs.\ (\ref{eq:fTheta}) and 
(\ref{eq:defTheta}).
As discussed in the text, for helium only two such functions should be used. 
\label{fig:correl}}
\end{figure}

In figure \ref{fig:correl} we show the correlation functions for the
four $p$-shell nuclei. Apart from the case of $^8$Be the central
scalar correlation function changes little from the SI to SD calculation.
Indeed, one
could almost keep this one fixed and only vary the additional
three. This may be due to some kind of perturbation argument: in
general the spin- and/or isospin-dependent correlation functions are a
lot smaller than the central one. The spin and isospin 
correlation 
functions are almost equal in magnitude but opposite in sign. There is
no obvious argument why this should be so, but it may be due to some
approximate persistence of the corresponding exact relation in the
case of $^4$He.

\begin{table}[htb]
\begin{center}
\caption{A comparison of the binding energies (in MeV) of
$^4$He and $^{16}$O using various techniques with the Malfliet-Tjon 
MT V  potential. TICI2 is the method presented in 
this paper, 
VMC are various variational Monte Carlo calculations, 
GFMC and DMC are fixed-node Green's function and diffusion Monte Carlo 
methods, FHNC is the  Fermi hypernetted chain method based on a Jastrow 
wave function, IDEA is the so-called integro-differential equation approach, 
and SVM is the stochastic variational method.}
\label{tab:MTV}
\begin{tabular}{lll}
& $^{4}$He & $^{16}$O \\
\hline
TICI2(SI)& 29.45 & 966.65 \\
TICI2(SD)& 29.45 & 973.67 \\
VMC& & $1024\pm5$\ \cite{Zabolitsky}\\
   & & $1103\pm1$\ \cite{BG}\\
   & & $1138.5\pm 0.2$\ \cite{ChKr}\\
GFMC \cite{Zabolitsky}& $31.3\pm0.2$ & $1194\pm 20$\\
DMC  &$31.32\pm0.02$ \cite{BBFG92}&$1189\pm 1$ \cite{ChKr}\\
FHNC/0 \cite{CFFL92} & & 987 / 1152\\
FHNC//0 \cite{Krotscheck} & &1059 / 1055\\
IDEA  \cite{BFL95}&30.7--31.2&1021--1027\\
SVM \cite{VS} &31.360 \\
\hline
\end{tabular}
\end{center}
\end{table}

In order to be able to judge the power of our technology we compare 
several different methodologies for the purely scalar MT V potential in table 
\ref{tab:MTV}. For $^4$He our results compare favourably with the Green's 
function and diffusion Monte Carlo results, both of which have now 
been superseded by various highly 
accurate calculations, including one based on  the stochastic variational 
principle (see ref.\ \cite{VS} and references in that paper).
 More surprising is the comparison for 
$^{16}$O. For the MT V potential, with its extremely strong binding, $^{16}$O 
is a high-density system. Our calculations with only two-body correlations are 
able to reproduce over 80\% of the binding energy (974 MeV for TICI2(SD) as 
compared to 1189 MeV for the very accurate DMC result). Inclusion of 
state-independent three-body correlations in our calculation -- 
a relatively straightforward extension -- should  probably be able to 
give a highly accurate answer. 

The TICI2 method also appears to be competitive with several of the methods
based on Jastrow-correlated non-interacting wave functions of a single Slater
determinant form, where the Jastrow correlation operator is a product over all
pairs of particles of a scalar (state-independent) pair function of
the interparticle distance only. The many-body energy expectation value for 
such trial Jastrow wave functions cannot be calculated analytically, but may
be calculated within controlled error bands using variational Monte Carlo
(VMC) techniques. Each of the three VMC results cited in table \ref{tab:MTV}
is of this form, with the differences between them resulting from different
choices of single-particle wave functions and from different forms of the 
Jastrow correlation function over which the variational search is made. 
These are discussed in more detail below. As an alternative to the 
stochastically exact VMC calculations of the energy expectation value, the
various Fermi hypernetted chain (FHNC) approximations 
\cite{ACFF96,CFFL92,Krotscheck} represent massively resummed cluster expansions
of the same quantity, in which some cluster terms (of chain or ladder form)
are summed to all orders but various so-called elementary diagrams are
neglected or approximated. The resulting FHNC approximants for the ground-state
energy are no longer strict variational bounds, insofar as the higher-order
terms have been neglected.

Firstly, the VMC result of $1024\pm5$ MeV for the binding of $^{16}$O
with the MT V potential from ref.\ \cite{Zabolitsky} uses
single-particle wave functions generated by a Woods-Saxon potential
and a scalar Jastrow correlation function obtained from a constrained
Euler-Lagrange optimisation of the second order cluster expansion
approximation to the energy functional.  We note that the
corresponding result from the lowest order FHNC calculation, in the
formulation of Fantoni and Rosati, the so-called FHNC/0 method
\cite{CFFL92} in which the elementary diagrams are neglected completely, 
using the same single-particle wave functions and Jastrow correlation
function, is $987$ MeV. Secondly, the VMC result of $1103\pm1$ MeV
from ref.\ \cite{BG} for the $^{16}$O binding energy employs harmonic
oscillator single-particle wave functions and a Jastrow correlation
function of Gaussian form, $f(r)=1+a\exp(-r^2/b^2)$, with the
oscillator length parameter and the constants $a$ and $b$
optimised. Finally, the similar VMC result of $1138.5\pm 0.2$\ from
ref.\ \cite{ChKr} uses optimised harmonic oscillator single-particle
wave functions and a Jastrow correlation function of the form
$f(r)=\exp(-a \e^{-br})$.

We note that a so-called FHNC//0 calculation for $^{16}$O using the MT
V potential has been performed by Krotscheck \cite{Krotscheck}. By
using the best possible choices for both the single particle wave
functions and the Jastrow correlation function (i.e., by solving the
corresponding Euler-Lagrange equation which arise by making the
FHNC//0 energy functional stationary) he obtains a value of 1059 MeV
for the binding energy. Furthermore, by optimising only the Jastrow
correlation function via his Euler-Lagrange approach, and using
single-particle wave functions obtained from a Woods-Saxon potential
which reproduces the r.m.s.\ radius of the GFMC calculation of ref.\
\cite{Zabolitsky}, he obtains the only very slightly worse value of
1055 MeV.  He estimates, using the FHNC/$C$ approximation that the
contribution from the neglected exchange diagrams is $14$ MeV.  We
also note that by employing precisely the same nearly optimal
Woods-Saxon single-particle wave functions, but with a Jastrow
correlation function obtained from a constrained minimisation of the
two-body cluster expansion approximation to the energy functional, a
corresponding FHNC/0 calculation \cite{CFFL92} gives a $^{16}$O
binding energy of $1152$ MeV. This difference of nearly $100$ MeV is
presumably a measure of the uncertainty in the FHNC/0 results, in the
light of the other results cited. It seems rather unlikely that the
corresponding VMC calculation for this same wave function will give
such a large value for the binding energy.  It would thus appear that
the latter FHNC/0 result does not provide a true variational
estimate. We defer to sec.\ \ref{sec:DisOut} a further discussion of
the FHNC results.

\begin{table}[htb]
\begin{center}
\caption{A comparison of the binding energies (in MeV) of
$^{12}$C and $^{16}$O using various techniques with the Brink-Boeker B1 
potential. For a description of the methods see the caption of the 
previous table. FAHT is a cluster expansion.} \label{tab:B1}
\begin{tabular}{lll}
& $^{12}$C & $^{16}$O \\
\hline
TICI2(SI)& 84.91 & 145.94 \\
TICI2(SD)& 103.93 &  167.30\\
VMC \cite{BBG}& $82.9\pm0.2$& $150.9\pm0.3$\\
FHNC/0 && 168.2$^a$ / 167.2$^b$ \cite{CFFL92}\\
       &68.9$^c$\cite{ACFF96} & 161.7 \cite{ACFF96}\\
FHNC-1 
       &       & 150.4$^a$ / 152.4$^b$ \cite{CFFL92}\\
      & 63.7$^c$ \cite{ACFF96} & 151.4$^c$ \cite{ACFF96} \\
FAHT-4  \cite{BBG} & 82.3 & 148.6 \\
IDEA   \cite{BFL95} & 79.8--80.1 & 164.4--165.2\\
\hline
\multicolumn{3}{l} {\small $^a$ state-independent Gaussian, $LS$ coupling}\\
\multicolumn{3}{l} {\small $^b$ state-independent 2-body Euler-Lagrange, $LS$ coupling}\\
\multicolumn{3}{l} {\small $^c$ state-independent 2-body Euler-Lagrange, $jj$ coupling}\\
\end{tabular}
\end{center}
\end{table}

The reason for investigating the MT V potential is that quite a few 
results are available, including very good GFMC and DMC 
calculations. However, we also investigate the B1 and MS3 potentials which 
have more sensible saturation properties, but for which no ``exact'' results
are available to act as a benchmark for the calculations. 
 B1 is a Wigner-plus-Majorana potential, 
whereas MS3 is a full V4-type potential. It is interesting to look at 
both potentials in turn, since they highlight different aspects of
the use of V4-type correlations. 

For the Brink-Boeker B1 potential we present results in table \ref{tab:B1}
for the nuclei $^{12}$C and 
$^{16}$O, the only nuclei for which several other results are available.
For $^{12}$C we must distinguish between calculations like our own that studied
the intrinsic state for the ground-state band and those that start from a
spherical state. Unfortunately, there are no GFMC or DMC results available
for either nucleus with the B1 potential, and we can only compare with the
results from other many-body techniques. All of the other results in table
\ref{tab:B1}, except those from IDEA, are based on a Jastrow form
of the wave function, again using a scalar (state-independent) correlation
operator. The results labelled FAHT-$n$ are from an $n$th order cluster
expansion of the Jastrow energy functional in the so-called factorised
Aviles-Hartogh-Tolhoek decomposition \cite{AHT}, which is multiplicative
van Kampen-like version of the original additive Ursell-like AHT version.

Our SI calculation for $^{12}$C results in
slightly better values for the
binding energy than the VMC and FAHT calculations. Note,
however, that Jastrow calculations with anisotropic HO wave functions
result in an increase of the binding energy of around 10 MeV
($92.3  \pm 0.3 \ {\rm MeV}$  with the VMC method of \cite{BBG}).
However, when
we include state-dependent (SD) correlations in our TICI2 calculation we
find an increase in binding of nearly 20 MeV. Bearing in mind that
our results are fully variational it is clear that our TICI2(SD)
calculation is better than any of the other results shown for $^{12}$C with 
the B1 potential. We note also that the results based on a spherical
assumption give appreciably less binding for $^{12}$C. This is particularly
so for the FHNC results \cite{ACFF96} based on a $jj$ coupling scheme.
We show results for both the lowest-order FHNC/0 approximation of
the Fantoni-Rosati scheme, in which all elementary diagrams are neglected,
and for their FHNC-1 approximation in which the first-order elementary
exchange diagram has been included. Such diagrams are clearly likely to
be of more importance for potentials of the B1 type with an appreciable
Majorana component than for purely scalar potentials of the MT V type.
The effect of their inclusion is seen to reduce the binding. The 
discrepancy between the VMC 
result \cite{BBG} of $82.9\pm0.2$ MeV
and the FHNC-1 result  \cite{ACFF96} in the $jj$ basis of
63.7 MeV is surprisingly large. We note, however, that the
state-independent Jastrow calculations in ref.\ \cite{ACFF96} were
obtained in a calculation which included the Coulomb force, by 
minimisation of the energy at the second order of the cluster expansion.
The results shown in table \ref{tab:B1} have been obtained by
subtracting out the quoted values of the Coulomb energy to the binding
energies given in ref.\ \cite{ACFF96}. Clearly, it is conceivable
that had the minimisation been done without the inclusion of the
Coulomb force, slightly improved results might have been obtained.
We also note that the calculations in ref.\ \cite{ACFF96} were performed
with Woods-Saxon single-particle wave functions, thereby making it impossible 
to remove centre-of-mass effects exactly.

In the case of $^{16}$O with the B1 potential we find
similar results to the $^{12}$C case. Again our TICI2(SI) calculation
gives a good result, and the improvement obtained by allowing for 
state-dependent correlations leads to what is possibly the best
(and certainly the variationally most reliable) result now available for this
case. The difference in binding between our SI and SD calculations of
over 20 MeV is clearly very appreciable. We note that the FHNC/0 and
FHNC-1 results for $^{16}$O have been done in both $LS$ \cite{CFFL92} and $jj$ 
\cite{ACFF96}  coupling schemes. For the former case we cite results using
Jastrow correlation functions both parametrised as a Gaussian form,
$f(r)=1+a\exp(-r^2/b^2)$, and from a constrained Euler-Lagrange
minimisation of the second-order cluster energy. Woods-Saxon
single-particle wave functions were used for all the FHNC results cited.

\begin{table}[htb]
\begin{center}
\caption{A comparison of the binding energies (in MeV) of
$^{12}$C and $^{16}$O using various techniques with the MS3 
potential. For a description of methods see the caption
of tables \protect{\ref{tab:MTV}} and \protect{\ref{tab:B1}}.
BHF is the Brueckner-Hartree-Fock method.} \label{tab:MS3}
\begin{tabular}{lll}
& $^{12}$C & $^{16}$O \\
\hline
TICI2(SI)& 46.22 & 85.56 \\
TICI2(SD)& 62.99 & 105.64\\
FHNC/0 &      & 113.5$^a$ \cite{CFFL92} \\
       & 36.5$^b$/54.2$^c$ \cite{ACFF96} & 108.1$^b$/145.5$^c$ \cite{ACFF96}\\
FHNC-1 &      & 105.3$^a$ \cite{CFFL92} \\
       & 34.7$^b$ \cite{ACFF96} & 103.0$^b$ \cite{ACFF96}\\
FAHT-3 (deformed) \cite{GFMP} & 51.4 & 107.7\\
BHF \cite{GFMP} & & 118.6\\
IDEA  \cite{BFL95} & 44.2--44.4
& 102.9--103.2\\
\hline
\multicolumn{3}{l} {\small $^a$ state-independent 2-body Euler-Lagrange, $LS$ coupling}\\
\multicolumn{3}{l} \small {$^b$ state-independent 2-body Euler-Lagrange, $jj$ coupling}\\
\multicolumn{3}{l} {\small $^c$ isospin-dependent 2-body Euler-Lagrange, $jj$ coupling}\\
\end{tabular}
\end{center}
\end{table}

We turn lastly to the case of the MS3 potential, and we again compare
our results with others for this potential for the same two nuclei,
$^{12}$C and $^{16}$O. We note that results for these nuclei for the
MS3 potential are also available within the FHNC framework using the
$jj$ coupling scheme and allowing, for the first time, some
state-dependence in the Jastrow correlation operator. More
explicitly, the authors of ref.\ \cite{ACFF96} incorporate 
isospin-dependence by allowing different pairwise Jastrow functions for the
proton-proton, neutron-neutron and proton-neutron pairs. Otherwise the
FHNC results shown in table \ref{tab:MS3} are calculated as explained
above for the B1 potential. We note again that the calculations of
ref.\ \cite{ACFF96} have included the Coulomb force. The results in
table \ref{tab:MS3} have been obtained by subtracting out the cited
Coulomb energy in each case. Once again, for $^{12}$C our
results are better than all the others. The effect of the isospin-dependence
in the FHNC calculation is comparable in magnitude to the effect of
the state-dependence in our calculations. However, the FHNC results
\cite{ACFF96} are less bound for $^{12}$C than our TICI2 results, presumably
at least in part because of their use of a spherical reference state.

For the case of $^{16}$O nucleus with the MS3 potential, the effect of
the incorporation of the state-dependent correlations in our TICI2
calculations is again to increase the binding energy by about 20
MeV. Naively, one might have expected a somewhat larger increase than
for the B1 potential, since the MS3 potential exploits the full V4
state dependence. Our own TICI2(SD) results are in this case close to
most of the other results cited, with the exception of the
isospin-dependent FHNC/0 result, where the gain in binding over the
corresponding state-independent case is nearly 40 MeV. In view of the
non-variational nature of both the FHNC and Brueckner-Hartree-Fock
(BHF) results, it is difficult to draw more detailed conclusions.

\section{Discussion and outlook\label{sec:DisOut}}

We have demonstrated that the TICI2 methodology provides a very
reasonable starting point for the calculation of the binding energies
of light-to-medium mass nuclei. The important state-dependent pairwise
correlation effects can be efficiently incorporated without destroying
the variational bound. By contrast, most competing many-body methods
suffer from various uncontrolled approximations, which often include
the loss of a strict upper bound on the ground-state energy, even when the
calculation starts from a variational framework. Such is often the case 
for FHNC and other more ambitious calculations performed within the 
correlated basis function (CBF) approach. Since such techniques provide
perhaps the main competitor to our own approach outlined here, it is worth
describing in some detail the source of the uncertainties that can
typically arise in FHNC-type calculations.

In a nutshell, the standard FHNC formulation of Fantoni and Rosati
sums all of the Jastrow cluster-expansion terms that can readily be
summed by one-dimensional integral equations (see ref.\ \cite{ACFF96}
and references cited therein for details). One drawback of this
procedure is that certain cancellations between elementary exchange
diagrams are thereby ignored. The resulting FHNC/$n$ scheme then
classifies the only omitted diagrams, the so-called elementary diagrams,
according to their number of vertices or points. Thus, the FHNC/$n$
approximation includes elementary diagrams with up to $n$ vertices. The
resulting prescription is both clear-cut and leads to relatively simple
integral equations which are valid for small distances. Nevertheless,
the long-range parts are more problematic. In particular, for infinite
homogeneous systems one neither expects to get nor obtains the correct
behaviour of the liquid structure function $S(k)$ as $k\rightarrow 0$.
Moreover this prescription does not allow, in principle, for correct
optimisation.

In contrast to the Fantoni-Rosati FHNC/$n$ scheme, Krotscheck has
proposed a different classification which maintains the Pauli
principle at each step.  One possibility among several to accomplish
this is to classify the exchange diagrams with respect to the number
of internal lines rather than vertices. The resulting lowest-order
approximant in this scheme is termed FHNC//0. In some well-defined
sense it sums the union of the equivalents of the ring diagrams with a
bare particle propagator and bosonic ladder terms. Effectively, for
the case of infinite homogeneous systems, it sums correctly and
self-consistently the most important diagrams in coordinate space when
short-range correlations are important, and in momentum space when
long-range correlations are important. In turn this leads to an
inconsistency between the pair distribution function, $g(r)$, and the
liquid structure function, $S(k)$. 

{}From the above brief discussion it should be clear that the main
difficulty and uncertainty in FHNC techniques is in the treatment of
elementary exchange diagrams. We have already seen in sec.\ \ref{sec:results}
the differences between the FHNC/0 and FHNC-1 results, where the latter
incorporates only the first-order elementary exchange diagram from the FHNC/4
set. We also note in this context that the FHNC/$C$ scheme of Krotscheck 
provides
another means of incorporating the omitted diagrams. Basically this scheme
exploits the fact that one knows the values of these diagrams in the case of
an infinite homogeneous system at momentum transfers $k=0$ and 
$k=2k_{\mathrm{F}}$, where $k_{\mathrm{F}}$ is the Fermi momentum. 
The FHNC/$C$ scheme
uses these exact properties to estimate the value of the omitted terms
by a simple polynomial interpolation between these limits. Thus
FHNC/$C$ is a superset of FHNC/0, but with an approximate treatment of what
the latter omits. Such a correction of the FHNC/0 equations is necessary for 
the optimisation. We have already commented in sec.\ \ref{sec:results} on the
discrepancy between the FHNC/0 and FHNC//0 results in the
case of $^{16}$O with the MT V potential. It has been shown \cite{Krotscheck}
that inclusion of the $C$ class of diagrams has only a marginal influence.

Apart from the above difficulties and uncertainties over the treatment
of elementary exchange diagrams, the incorporation of state-dependent
correlations into FHNC treatments is also very fraught with
difficulty.  This arises essentially due to the non-commutativity of
the correlation operators in a Jastrow-type product wave function when
the scalar Jastrow function is replaced by an operatorial sum. The
isospin-dependent correlations introduced in ref.\ \cite{ACFF96} very
recently, and reported in sec.\ \ref{sec:results}, have led to
workable FHNC calculations only because in this case the corresponding
operators commute among themselves. The introduction of more complicated
operatorial forms (e.g., of the sort used by us) will not provide such
a straightforward extension. Our own TICI2 method does not suffer from
these drawbacks.

We now turn briefly to the alternative IDEA approach, which is
formulated purely in terms of the hyper-radius and the relative
coordinate of the two particles needed to describe the pair
correlations incorporated in the method. In this method, as currently
practised, several uncontrolled and often unjustified approximations are
made. As a consequence the results are non-variational, and usually
one has a range of values depending on which of many approximations has
been made, none of which is fundamentally preferred to another. A
recent investigation \cite{rafa} of a simple model suggests rather
forcefully that even if {\em no} approximations are made, the
inclusion of the hyper-radial excitations in a many-body wave function
that lies at the heart of the IDEA method does not provide a very
cost-effective improvement. This study suggest that it would be much
more efficient to include pair (and triplet, etc.) correlations to the best
of one's ability, as done in the present work. From this viewpoint our own
method of incorporating pair correlations is far more systematic than
in the IDEA methodology.

In conclusion, our method works well for interactions and correlations of 
V4 form. Of
course one needs a much more complicated interaction to describe
phase shifts well: the latest Argonne V18 interaction has 18
components. Even though it may not be immediately obvious what kind of
correlations are important when one wishes to deal with such a
complicated force, it is clear that the first step is to look at
interactions and correlations of the V6 form, where one includes the
tensor force and tensor correlations. Such an investigation is now
under way, and it appears that the methodology sketched in
Sec. \ref{sec:method} can be extended to deal with such a more
realistic force. Actually when we have V6 type correlations, we may
already pick up the major part of the spin-orbit force as well; at
this level there is no easy match between the operatorial structure of
the correlations and that of the potential.

In the end we shall also have to include three-body
correlations. Since the effect of such correlations is not expected to
be very large, one might hope that inclusion of state-independent
three-particle correlations is enough. This would still lead to very
complicated calculations, but we believe that we can certainly deal with
such a scheme. Indeed, we have already reported such results for $^4$He
using purely central (scalar) potentials \cite{us}. Results for heavier nuclei
with more realistic potentials will be reported in the future.

\section*{Acknowledgements}
This work was supported by DGICyT (Spain) grant PB92-0820, and by a research
grant from the Engineering and Physical Sciences Research Council
(EPSRC) of Great Britain. One of us (P.I.M.) acknowledges DGICyT (Spain) 
for a fellowship.


\begin{thebibliography}{99}
\bibitem{carlson} N. S. Pudliner, V. R. Pandharipande, J. Carlson and 
R. B. Wiringa, Phys. Rev. Lett. {\bf 74} (1995) 4396.
\bibitem{gloeckle} W. Gl\"ockle and H. Kamada,
Phys. Rev. Lett. {\bf 71} (1993) 971.
\bibitem{rosati} A. Kievsky, M. Viviani and S. Rosati,
Nucl. Phys. A {\bf 551} (1993) 241; A {\bf 577} (1994) 511.
\bibitem{us} 
 R. F. Bishop, M. F. Flynn, M. C. Bosc\'a, E. Buend\'{\i}a and R. Guardiola,
Phys. Rev. C {\bf 42} (1990) 1341;  Condensed Matter Theories, vol. 5, 
V. C. Aguilera-Navarro, ed. (Plenum Press, N.Y., 1990), p. 253;
J. Phys. G: Nucl. Part. Phys. {\bf 16} (1990) L61;\\
 R. F. Bishop, E. Buend\'{\i}a, M. F. Flynn and R. Guardiola,
 Condensed Matter Theories, vol. 6, 
S. Fantoni and S. Rosati, eds. (Plenum Press, N.Y., 1991), p. 405;
J. Phys. G: Nucl. Part. Phys. {\bf 17} (1991) 857;
J. Phys. G: Nucl. Part. Phys. {\bf 18} (1992) 1157;
J. Phys. G: Nucl. Part. Phys. {\bf 19} (1993) 1663.
\bibitem{ACFF96}
F. Arias de Saavedra, G. Co', A. Fabrocini and S. Fantoni.
``Model calculations of doubly closed shell nuclei in CBF theory III. j-j
       coupling and isospin dependence''. Preprint nucl-th/9604013, Nucl. Phys.
A, in press.
\bibitem{VS} K. Varga and Y. Suzuki, Phys. Rev. C {\bf 52} (1995) 2885;
Phys. Rev. A {\bf 53} (1996) 1907.
\bibitem{kalos} J. Carlson and M. H. Kalos,
Phys. Rev. C {\bf 32} (1985) 2105.
\bibitem{pandha}
S. C. Pieper, R. B. Wiringa and V. R. Pandharipande,
Phys. Rev. C {\bf 46} (1992) 1741.
\bibitem{S3}
I. R. Afnan and Y. C. Tang, Phys. Rev. {\bf 175} (1968) 1337.
\bibitem{guardiola}
R. Guardiola, Nucl. Phys. A {\bf328} (1979) 490.
\bibitem{BrinkBoeker}
D. M. Brink and E. Boeker, Nucl. Phys. A {\bf 91} (1967) 1.
\bibitem{GFMP} R. Guardiola, A. Faessler, H. M\"uther and A. Polls, Nucl. 
Phys. A {\bf 371} (1981) 79.
\bibitem{MT} R. A. Malfliet and J. A. Tjon, Nucl. Phys. A {\bf 127} (1969) 161.
\bibitem{Zabolitsky}
J. G. Zabolitzky and M. H. Kalos, Nucl. Phys. A {\bf 356} (1981) 114;\\
U. Helmbrecht and J. G. Zabolitzky, Nucl. Phys. A {\bf 442} (1985) 109;\\
J. G. Zabolitzky, K.E. Schmidt and M.H. Kalos, 
Phys. Rev. C {\bf 25} (1982) 1111.
\bibitem{BG} E. Buend\'{\i}a and R. Guardiola, Condensed Matter Theories, vol. 8, 
L. Blum and F. B. Malik, eds. (Plenum Press, N.Y., 1993), p. 301.
\bibitem{ChKr} S. A. Chin and E. Krotscheck, Nucl. Phys. A {\bf 560} (1993) 
151.
\bibitem{BBFG92}
 R. F. Bishop, E. Buend\'{\i}a, M. F. Flynn and R. Guardiola,
J. Phys. G: Nucl. Part. Phys. {\bf 18} (1992) L21.
\bibitem{CFFL92} 
G. Co', A. Fabrocini, S. Fantoni and I. E. Lagaris, Nucl. Phys. A 
{\bf549} (1992) 439.
\bibitem{Krotscheck} E. Krotscheck, Nucl. Phys. A {\bf 465} (1987) 461.
\bibitem{BFL95} R. Brizzi, M. Fabre de la Ripelle and M. Lassaut, Nucl. 
Phys. A {\bf 596} (1996) 199.
\bibitem{BBG} M. C. Bosc\'{a}, E. Buend\'{\i}a and R. Guardiola, Phys. Lett. B {\bf198}
(1987) 312; Condensed Matter Theories,  vol. 3,
J. Arponen, R. Bishop, M. Manninen, eds. 
(Plenum Press, N.Y., 1987), p. 101.
\bibitem{AHT}
J. B. Aviles, Jr., Ann. Phys. (NY) {\bf 5} (1958) 251;\\
C. D. Hartogh and H. A. Tolhoek, Physica {\bf 24} (1958) 721, 875, 896;\\
J.W. Clark and P. Westhaus, J. Math. Phys. {\bf 9} (1968) 131.
\bibitem{rafa} R. Guardiola, P. I. Moliner and J. Navarro
``Two-body and Hyperradial Correlations in the description of Many-Body
Systems'', Phys. Lett. B, in press.
\end{thebibliography}
\end{document}